\documentclass[12pt]{article}
\usepackage{authblk}
\usepackage{geometry}
\usepackage{amsmath}
\usepackage{amssymb}
\usepackage{amsthm}
\usepackage{mathrsfs}
\usepackage{subcaption}
\usepackage{customcommands}
\usepackage{bm}
\usepackage{Nettastyle}

\usepackage{physics} 
\usepackage{tensor} 
\usepackage{tcolorbox}

\usepackage[normalem]{ulem}

\preprint{MIT-CTP/5371}

\title{Negative Complexity of Formation: the Compact Dimensions Strike Back}

\author{Netta Engelhardt}
\author{and \AA{}smund Folkestad}
\emailAdd{engeln@mit.edu}
\emailAdd{afolkest@mit.edu}
\affiliation{Center for Theoretical Physics, Massachusetts Institute of Technology, \\Cambridge, MA 02139, USA}

\abstract{
We show that the vacuum-subtracted maximal volume, the proposed holographic dual to complexity of formation,
can be negative when contributions from compact directions are included. We construct explicit solutions with
arbitrarily negative complexity of formation in asymptotically AdS$_{4}\times S^{7}$
SUGRA. These 
examples rely critically on the compact directions, specifically the fact that the full eleven-dimensional spacetime is not asymptotically AdS$_{11}$. While
there is some ambiguity in the extension of the holographic complexity proposal to the compact directions, we show that
the two natural candidates can both have arbitrarily negative complexity of formation in SUGRA solutions.
We further find examples in which complexity can even \textit{decrease}
at late times, including cases of both single-sided geometries and two-sided 
wormholes. In particular, we construct a cosmological wormhole
with simultaneously negative and decreasing complexity of formation (as computed by volume) at late times.
We find a distinguished role for relevant primaries in these constructions and comment on possible interpretations.
}

\begin{document}

\maketitle

\section{\label{sec:intro}Introduction}

Recent progress on the emergence of spacetime has crucially relied on the
geometrization of quantum information theoretic quantities
\cite{RyuTak06,RyuTak06-2,HubRan07,Wal12,CzeKar12,FauLew13,LewMal13,Don13, MalSus13, EngWal14,Sus14a,StaSus14,BroRob15,BroRob15b, JafLew15,Don16}.
A relative newcomer to this set of connections has been the geometrization of computational complexity 
\cite{Sus14a,StaSus14,BroRob15,BroRob15b}, either through the proposed Complexity=Action \cite{BroRob15,BroRob15b} or 
Complexity=Volume duality \cite{StaSus14}. The latter, which is our focus in this article, relates the circuit complexity
$\mathcal{C}$ of a given holographic 
CFT$_{d}$ state $\ket{\psi(\tau)}$ relative to some reference state
$\ket{R}$ to regulated bulk spatial volumes: 
\begin{equation} \begin{aligned}\label{eq:CVdef}
    \mathcal{C}\big(\ket{\psi(\tau)}, \ket{R}\big) = \max_{\Sigma} \frac{
        \vol[\Sigma] }{ G_N L } \equiv
    \mathcal{C}_V\left(\ket{\psi(\tau)}\right),
\end{aligned} 
\end{equation} 
where $\Sigma$ a bulk hypersurface that intersects the conformal 
boundary on the timeslice $\tau$;  $L$ is a length scale which we will take to be the AdS
radius, and $\mathcal{C}_V$ a convenient shorthand for the gravitational
quantity. 

In a recent paper \cite{EngFol21a}, we proved that the \textit{complexity of formation}
$\mathcal{C}_F$ \cite{BroRob15b, ChaMar16} satisfies
\begin{equation}\label{eq:PCT}
\begin{aligned}
    \mathcal{C}_F(\ket{\psi}) \equiv \mathcal{C}_V\left(\ket{\psi}\right) -
    \mathcal{C}_V\left(\ket{0}\right) \geq 0,
\end{aligned}
\end{equation}
with equality if and only if $\ket{\psi}=\ket{0}$, where $\ket{0}$ is the vacuum
dual to pure AdS$_{d+1}$.  We established this for asymptotically AdS$_{d+1}$ spacetimes under the assumption of the weak curvature condition
(WCC):
\begin{equation}\label{eq:WCC} 
    t^a t^b \left(R_{ab}-\frac{ 1 }{ 2 }g_{ab}R - \frac{ d(d-1) }{ L^2 }g_{ab} \right)
    \geq 0, \qquad \forall \text{\ timelike } t^a,
\end{equation} 
which in Einstein gravity reduces to the
weak energy condition (WEC): $T_{ab}t^a t^b \geq 0$ for all timelike $t^{a}$. 
The result \eqref{eq:PCT} implies that among
the states with a classical asymptotically AdS$_{d+1}$ dual respecting the WCC, the vacuum is
the least $\mathcal{C}_V$--complex. That is, WCC-respecting excitations of the vacuum
move away from the reference state as measured by the complexity. The necessity of the WCC is clear from \cite{ChaGe19, BerGal20}, who found examples in which the vacuum-subtracted volume is negative; in those examples the WCC is violated.

The assumption of the WCC is somewhat unnatural from a holographic perspective: consistency conditions in the large-$N$,
large-$\lambda$ limit of the AdS/CFT correspondence are typically proven using the Null Curvature Condition (NCC)
$R_{ab}k^{a}k^{b}\geq 0$ for null vectors $k^{a}$. The latter (strictly weaker) condition is expected to be true for any valid
classical matter; the same, however, is not true for the WCC~\cite{HerHor03}. Nevertheless, it turns out that the WCC holds
in type II and eleven-dimensional SUGRA (see appendix~\ref{sec:WECsupergravity}): even though the dimensional reduction of an
asymptotically AdS$_{d+1}\times K$ over the compact dimensions $K$ may violate the WCC while satisfying the NCC, inclusion
of the compact dimensions restores the WCC. Prima facie, then, it may be tempting to conclude that when working
in full ten or eleven dimensional SUGRA, our results immediately imply that complexity of formation is
\textit{always} positive. It would then be natural to conclude that $\mathcal{C}(\ket{\psi}, \ket{R})$ should be identified with $\mathcal{C}_F(\ket{\psi})$, with a reference state $\ket{R}$ that is identically the vacuum $\ket{0}$.

This naive conclusion, however, suffers from several flaws. First, the Complexity=Volume proposal does not admit an
obvious generalization allowing the inclusion of compact directions. There are (at least) two natural candidates: (1) the volume of
the maximal volume slice $\Sigma_{\mathrm{full}}$ in the full AdS$_{d+1}\times K$ spacetime; or (2) the volume of the
maximal volume slice $\Sigma_{\mathrm{reduced}}$ extended in the non-compact
directions. It is simple (see
Sec.~\ref{sec:SUGRA}) to show that in general the dimensional reduction of $\Sigma_{\rm full}$ does not result in $\Sigma_{\rm
reduced}$, and that consequently
$$\vol[\Sigma_{\mathrm{full}}]\neq \vol[\Sigma_{\rm reduced} \times K]. $$
A maximal volume slice in the full spacetime need not be maximal in the AdS directions, and vice versa. 

Our goal here, however, will not be to argue in favor of either (1) or (2) but to demonstrate that \textit{neither}
candidate can avoid a negative complexity of formation: the WCC restoration that accompanies the inclusion of compact
directions fails to save either candidate from predicting that certain valid
spacetimes in AdS/CFT are simpler
than the vacuum\footnote{By `valid' here we mean a stricter definition than is typically used (which is often just the requirement of the NEC and global hyperbolicity: here we mean that they are inherited from top down truncations of SUGRA.}. As a consequence,
since $\mathcal{C}_F$ is not positive semi-definite, it cannot be reinterpreted
as the complexity with the vacuum as the reference state. To simplify matters, we will demonstrate this in a special case where the two candidate proposals
coincide: a moment of time symmetry. In such a case, $\Sigma_{\mathrm{full}}$  reduces in the AdS directions to
$\Sigma_{\mathrm{reduced}}$, so that any conclusions are free of ambiguities
relating to a choice between (1) and (2). 

This result may at first appear to contradict our proof in~\cite{EngFol21a}. How can the vacuum-subtracted maximal volume be
negative in spacetimes respecting the WCC? The answer is a prime realization of the principle of conservation of misery:
while the inclusion of compact directions restores the WCC, it in turn violates our assumption about AdS$_{d+1}$
asymptotics. Thus we have two (mutually exclusive) options: accept violations of
WCC in the absence of compact directions, or accept violations of AdS$_{d+1}$ asymptotics. As it turns out, either option leads to states with negative complexity of formation. 

In the following, we find a family of AdS$_4$ initial data supported by scalar tachyons above the
Breitenlohner-Freedman bound~\cite{BreFre82}, inherited from a truncation and dimensional reduction over the compact directions
of eleven-dimensional SUGRA. In the full asymptotically AdS$_{4}\times S^{7}$ data, the WCC is satisfied, but
upon dimensional reduction the resulting spacetimes violate the WCC and satisfy
the NCC. These geometries come in two flavors~\footnote{For computational facility, the examples with boundary sources are
not constructed in an exact dimensional
reduction of eleven-dimensional SUGRA, but instead with a slightly
modified (but qualitatively similar) scalar potential, which enables analytical solutions. Analogous
one-sided spacetimes were considered in $D=11$ SUGRA reduced to AdS$_4 \times
S^7$
in \cite{HerHor04,HerHor05}, and we expect all of our qualitative findings in
Sec.~\ref{sec:vacdecay} to apply in $D=11$ SUGRA.}: with and without boundary sources. For the former, the inclusion of boundary sources yields asymptotically AdS$_4$ spacetimes undergoing AdS false vacuum decay; such spacetimes, again supported by scalar tachyons, have a negatively divergent complexity of formation that decreases at late times. Among these spacetimes is a novel
cosmological wormhole; even though the spacetime connects two asymptotic
boundaries (with no dS region in between~\cite{FisMar14}), we find that the holographic volume
complexity is nevertheless \textit{smaller} than that of pure AdS. 
While spacetimes with negative and divergent $\mathcal{C}_F$ due to boundary sources 
were previously considered by \cite{Moo17b}, our examples without boundary sources are quite distinct, and should be
regarded as the main finding of this paper (although we also expand on examples with boundary sources, more analogous to the ones discussed in~\cite{Moo17b}).
In this case we find initial 
data with arbitrarily negative (but finite) complexity of formation, both when viewed in eleven and four dimensions.
The arbitrarily low complexity in this case is not caused by altering the boundary theory. Instead, it 
is obtained by a smooth deformation of the CFT state away from vacuum, and the low $\mathcal{C}_V$ is a genuine IR-effect caused by the compact dimensions.

There is however an
underlying common denominator to all of our examples: 
they are constructed by turning on relevant scalar primaries in the CFT. 
This pattern together with the theorems of \cite{EngFol21a} 
suggests potential insights into the landscape of low
complexity holographic states. If tachyonic bulk scalars, which are dual to
relevant CFT scalar primaries, happen to be the only WEC-violating fields, then the only way to reduce the ``distance'' -- as measured by
$\mathcal{C}_V$ -- to $\ket{R}$ below
 the fixed nonzero value $\mathcal{C}_V(\ket{0})$ is to turn on VEVs for
 relevant scalars. Other operators will be dual to WEC-respecting fields, and so
 turning them on will cause $\mathcal{C}_V$ to increase with respect to the vacuum
 value.  Thus, the presence of unstable directions of the IR fixed point correlates with the
possibility of reducing complexity below the vacuum value. This could be due to the fact that the vacuum of a
potential gapped phase at the end of the RG flow has significantly fewer
correlations, simplifying the preparation of the state.

Our particular examples of low complexity spacetimes also provide potential
insight into the reference state $\ket{R}$ implicit in the CV
proposal. These examples are 
constructed by creating pockets of approximately constant scalar field at a moment of time-symmetry in the bulk, resulting in an effective AdS radius smaller than the asymptotic value $L$ within the pocket.  We find that $\mathcal{C}_F$ becomes arbitrarily negative
as the pocket becomes larger: that is, the complexity becomes progressively closer to that of $\ket{R}$ via this reduction of the effective AdS radius in an increasingly
large region. Since the limit of small AdS radius is not a well-defined classical geometry, this finding is 
consistent with the common perspective that $\ket{R}$ is a state without a geometric dual, e.g. a set of factorized
qubits. 

What is the upshot of our results for the Complexity=Volume proposal? At
minimum, there is need for an unambiguous prescription that accounts for
contributions from compact dimensions. It is clear that volumes can have a
qualitatively different behavior when compact dimensions are included. On a more
speculative level, there appears to be a sharp distinction (in the
dimensionally-reduced picture) between operators whose dual is WCC-respecting
and WCC-violating; acting on the vacuum with the former can only increase
complexity; the latter, however, can decrease complexity. It would be
interesting to understand this better in the dual CFT, perhaps using the
proposed definitions of complexity in~\cite{BelLew18, ChaCha21,FloHel20a,FloHel20b}. 

The paper is structured as follows. In Sec.~\ref{sec:SUGRA} we present our SUGRA maximal volume asymptotically AdS$_4 \times S^7$
initial data, together with its dimensional reduction. Then, to ensure that our constructed initial data gives the
unique maximal volume slice in the evolved spacetime, we derive general properties of maximal volume slices in type
II and $D=11$ SUGRA in Sec.~\ref{sec:generalProps}. Finally, in Sec.~\ref{sec:vacdecay} we turn on boundary sources to study spacetimes undergoing AdS vacuum decay, both one-sided
and two-sided. The appendix~\ref{sec:appendix} provides technical details omitted in the main text.

\section{Lower Unbounded $\mathcal{C}_F$ in SUGRA}\label{sec:SUGRA} We begin by constructing asymptotically AdS$_{4}\times S^{7}$ examples with negative complexity of formation supported by a well-studied truncation of eleven-dimensional SUGRA compactified on the $S^7$
\cite{CveDuf99}\footnote{ 
This theory was used to construct big crunch geometries in \cite{HerHor04,
HerHor05}. Our boundary conditions will differ from \cite{HerHor04, HerHor05},
so that the boundary dual will be different. The spacetimes in the next section will
more closely resemble the situation in  \cite{HerHor04, HerHor05}.
}
\begin{equation}
    \begin{aligned}\label{eq:4Dtheory}
    S &= \frac{ 1 }{ 8\pi G_N }\int_{M}\dd^{4}x \sqrt{-g} \left[\frac{ 1 }{ 2 }R +
        \frac{ 3 }{ L^2 } - \frac{ 1 }{ 2 } (\nabla \phi)^2  - V(\phi)\right],
\end{aligned}
\end{equation}
with scalar potential 
\begin{equation}
\begin{aligned}
    V(\phi) = \frac{ 1 }{ L^2 }\left(1 - \cosh \sqrt{2}\phi \right).
\end{aligned}
\end{equation}
Since $V(\phi)$ is unbounded below, this theory violates the WEC (and
equivalently the WCC). However, it is simple to check that the tachyonic scalar
mass about the $\phi=0$ vacuum is above the BF bound. Furthermore, the null
energy condition is satisfied, as always is the case for minimally coupled scalars.

A solution to the equations of motion of \eqref{eq:4Dtheory} with
four-dimensional line element $ds^{2}_{4}$ lifts to a solution of eleven
dimensional SUGRA with geometry 
\begin{equation}
\begin{aligned}
    \dd s^2 = \Delta^{2/3} \dd s_4^2 + \frac{ 4L^2 }{ \Delta^{1/3} }\sum_{i=1}^{4} X_{i}^{-1}\left(\dd \mu_{i}^2 + \mu_{i}^2
    \dd \psi_i^2\right),
\end{aligned}
\end{equation}
where $\dd s_4^2$ is the four-dimensional metric and 
\begin{equation}
\begin{aligned}
    \mu_{i} &= \left(\sin\theta, \cos\theta \sin\varphi, \cos\theta \cos\varphi \sin \xi, \cos\theta \cos\varphi \cos
    \xi \right), \\
    X &\equiv  X_1 = X_2 =  e^{-\frac{ \phi }{ \sqrt{2} }},\\
    X_3 &= X_4 = X^{-1}, \\
    \Delta &= \sum_{i=1}^{4} X_{i}\mu_{i}^2.
\end{aligned}
\end{equation}
The angles $(\theta, \varphi, \psi_1, \psi_2, \psi_3, \psi_4)$ run over the
range $[0, \pi]$, while $\xi \in [0, 2\pi)$.  If we set $\phi=0$ ($X_i=1$), then
the transverse space just becomes a round $S^7$ with radius $2L$: turning on the
scalar $\phi$ squashes the $S^7$. 

We now want to construct initial data on a spherically symmetric maximal volume slice
$\Sigma$ which has arbitrarily low complexity of formation in the $(d+1)=4$ theory
\eqref{eq:4Dtheory}. Furthermore, upon success of this endeavor, we want to investigate whether considering the
full volume in eleven-dimensional SUGRA restores positivity or boundedness from
below. Let us first note that in the spherically symmetric case, if $\Sigma$ has
embedding coordinates $(t(r), r, \Omega_{i})$ where $\Omega_{i}$ are the angles
on the $2$-sphere, then $\Sigma$ can be lifted to a slice $\tilde{\Sigma}$ in the eleven-dimensional spacetime with
embedding coordinates $(t(r), r, \Omega_i, \theta, \varphi, \xi, \psi_i)$. However,
the slice $\tilde{\Sigma}$ is generally not a maximal volume slice (even though $\Sigma$ is):
 turning on the scalar $\phi$ induces volume in the compact
dimensions, so if $\partial_t \phi\neq 0$ on $\Sigma$, then we can gain volume in
the eleven-dimensional spacetime $(\tilde{M}, \tilde{g})$ by deforming $\tilde{\Sigma}$. However, letting
$K_{ab}$ denote the extrinsic curvature, if we take a
moment of time symmetry, $K_{ab}[\Sigma]=\partial_t \phi=0$, then we will
also be at a moment of time-symmetry in eleven dimensions, and so
$\tilde{\Sigma}$ is also extremal in $(\tilde{M}, \tilde{g})$. 

In a moment we will construct explicit initial data, but let us first find an
expression for the volume of $\tilde{\Sigma}$. We can take our coordinates on
$\Sigma$ so that
\begin{equation}
\begin{aligned}
      \dd s^2 |_{\Sigma}   &   = B(r)\dd r^2 + r^2 \dd \Omega^2,   \\
    \dd s^2 |_{\tilde{\Sigma}} &= \Delta^{2/3}\left( B(r)\dd r^2 + r^2 \dd \Omega^2 \right) + \frac{ 4L^2 }{ \Delta^{1/3} }\sum_{i=1}^{4} X_{i}^{-1}\left(\dd \mu_{i}^2 + \mu_{i}^2
    \dd \psi_i^2\right),
\end{aligned}
\end{equation}
for some $B(r)>0$. Integrating out the compact dimensions, we find an effective
volume form $\bm{\tilde{\epsilon}}$ on $\Sigma$:\begin{equation}
\begin{aligned}
    \bm{\tilde{\epsilon}} = (2L)^7  f\left( e^{- \frac{ \phi }{ \sqrt{2}  }}\right) \bm{\epsilon},
\end{aligned}
\end{equation}
where $\bm{\epsilon}$ is the canonical volume form on $\Sigma$ induced in the
four-dimensional spacetime $(M, g)$ and
\begin{equation}
\begin{aligned}
    f(X)   &= \pi^4 \frac{ 9\left(1 + X^{2/3}\right)\left(2 + 4 X^{2/3} + 8 X^{4/3} + 7 X^2 + 8 X^{8/3} + 4 X^{10/3} + 2
    X^4\right) }{ 70 X^{1/3}(1 + X^{2/3} + X^{4/3})^3 }.
\end{aligned}
\end{equation}
When $\bm{\tilde{\epsilon}}$ is integrated over $\Sigma$, it gives the volume of
$\tilde{\Sigma}$. Since $f(X) \geq f(1)=\vol[S^{7}]$, including the compact
dimensions always increases the volume compared to the naive multiplication of
the $d=3$ volume with $(2L)^7 \vol[S^7]$. For large $|\phi|$, to leading order
\begin{equation}
\begin{aligned}
    \bm{\tilde{\epsilon}} = \frac{ 27 }{ 35 }(2L)^7 \vol[S^7] e^{\frac{ |\phi| }{ 3\sqrt{2} }}\bm{\epsilon}.
\end{aligned}
\end{equation}
Thus, for large scalar condensates there is generally an \textit{exponential}
difference in $|\phi|$ between the naive $(2L)^7 \vol[S^7] \vol[\Sigma]$ and the true
volume $\vol[\tilde{\Sigma}]$. This clearly demonstrates that compact directions
can dramatically modify the volume even if the extremal slice is unchanged.

We now want to pick initial data leveraging the negativity of $V(\phi)$ to
minimize the volume of $\Sigma$.  The solution of the constraint equations for
Einstein-Maxwell-Scalar theory on a spherically symmetric maximal volume slice
at a moment of time symmetry and with $d=3$ is \cite{EngFol21a}
\begin{equation}
\begin{aligned}
    \label{eq:omegad4}
    B(r) &= \left( 1+ r^2 - \frac{ \omega(r) }{ r }\right)^{-1}, \\
    \omega(r) &= \frac{ 1 }{ 2 }\int_{0}^{r}\dd \rho \rho^2 e^{\frac{ 1 }{ 2 }\int_{r}^{\rho}\dd z
    \phi'(z)^2}\left[\left(1+ \rho^2 \right)\phi'(\rho)^2 + 2 - 2 \cosh\sqrt{2}\phi \right],
\end{aligned}
\end{equation}
where we pick units of $L=1$ for brevity. The quantity $\omega(r)$ is a quasi-local mass function, and $\omega(\infty)$ is proportional to the
conserved spacetime mass when $\omega(\infty)$ is finite \cite{HerHor04}. 

We now pick the scalar profile on
$\Sigma$ to fall off in such a way so that the evolution of the initial data on
$\Sigma$ does not spoil the AdS asymptotics. Furthermore, we must ensure that
$B(r) > 0$ everywhere so that $\Sigma$ is everywhere spacelike. Beyond these two
constraints, we are free to choose the profile for
$\phi$.\footnote{We do not have a guarantee that it is possible to prepare
\eqref{eq:scalarprofile4D} via the Euclidean path integral.
However, for our conclusions to fail, it would have to be impossible to
construct any qualitatively similar scalar condensate at a moment of time symmetry, since our
findings do not depend on the particular quantitative details of the condensate \eqref{eq:scalarprofile4D}.
Any tachyonic scalar condensate should give the same conclusion as long as
we have (1) a pocket where $\phi$ is non-zero and 
approximately constant, (2) the pocket is of size at least $r\sim L$, and (3)
the scalar falls off as slowly as is consistent \eqref{eq:phiFalloffs} and finite energy.
It seems unlikely that such profiles cannot be prepared. In fact, \cite{MarPar17}
shows perturbatively that such profiles can be prepared using the Euclidean path integral. 
}
The usual near boundary analysis of the Einstein-Klein-Gordon system constrains
the asymptotic behavior of $\phi$: 
\begin{equation}\label{eq:phiFalloffs}
\begin{aligned}
    \phi(r, x) = r^{-\Delta_{-}}\left(\phi^{(0)}(x) +\phi^{(2)}(x) r^{-2} + \ldots\right) + r^{-\Delta_+}\left(\psi^{(0)}(x) +
    \psi^{(2)}(x) r^{-2} + \ldots\right),
\end{aligned}
\end{equation}
where $\Delta_{-}=1$ and $\Delta_{+}=2$. In order to avoid turning on boundary sources and to keep
$\omega(\infty$) finite (so the volume divergence structure agrees with that of pure AdS and so
the Balasubramanian-Kraus stress tensor is defined \cite{BalKra99b}), we take
$\phi^{(2n)}=0$. 

Let us construct one-sided initial data with no minimal surfaces, so that one
coordinate patch covers the whole of $\Sigma$ (this happens when $B(r)$ is
nowhere divergent \cite{EngFol21a}). We choose the profile 
\begin{equation}\label{eq:scalarprofile4D}
    \begin{aligned}
        \phi(r) = 1-e^{-a^2/r^2},
\end{aligned}
\end{equation}
which has the requisite $\mathcal{O}(r^{-2})$ behavior needed to keep
$\omega(\infty)$ finite. In the appendix we prove that in type II and $D=11$ SUGRA, (1) any $K=0$ slice is
maximal, and (2) there can only be one maximal slice at a fixed anchoring. Thus, the spacetime
obtained by evolving our initial data cannot possess another maximum volume slice with larger volume.

It can be checked that the profile~\eqref{eq:scalarprofile4D}
results in $0 < B(r) < \infty$,  so that our assumption of no minimal surfaces
is satisfied. We now proceed to calculate volumes of $\Sigma$ and
$\tilde{\Sigma}$ relative a constant$-t$ slice of AdS$_4$ and AdS$_4\times S^7$,
respectively:
\begin{equation}
\begin{aligned}
    \Delta V_{\Sigma}(a) = \vol[S^2] \int_{0}^{\infty}\dd r r^2 \left[\frac{ 1 }{ \sqrt{1 + r^2 - \frac{ \omega(r) }{
    r} }} - \frac{ 1 }{ \sqrt{1 + r^2 }} \right], \\
\Delta V_{\tilde{\Sigma}}(a) = \vol[S^2]2^7 \int_{0}^{\infty}\dd r r^2 \left[\frac{ f\left(e^{-\frac{ \phi }{
        \sqrt{2} }} \right) }{ \sqrt{1 + r^2 - \frac{ \omega(r) }{
        r} }} - \frac{ \vol[S^7] }{ \sqrt{1 + r^2 }} \right],
\end{aligned}
\end{equation}
(Here the factor $2^7$ appears since the round $S^7$ has radius $2$ in
units of $L=1$.)

\begin{figure}
     \centering
     \begin{subfigure}[b]{0.49\textwidth}
         \centering
         \includegraphics[width=\textwidth]{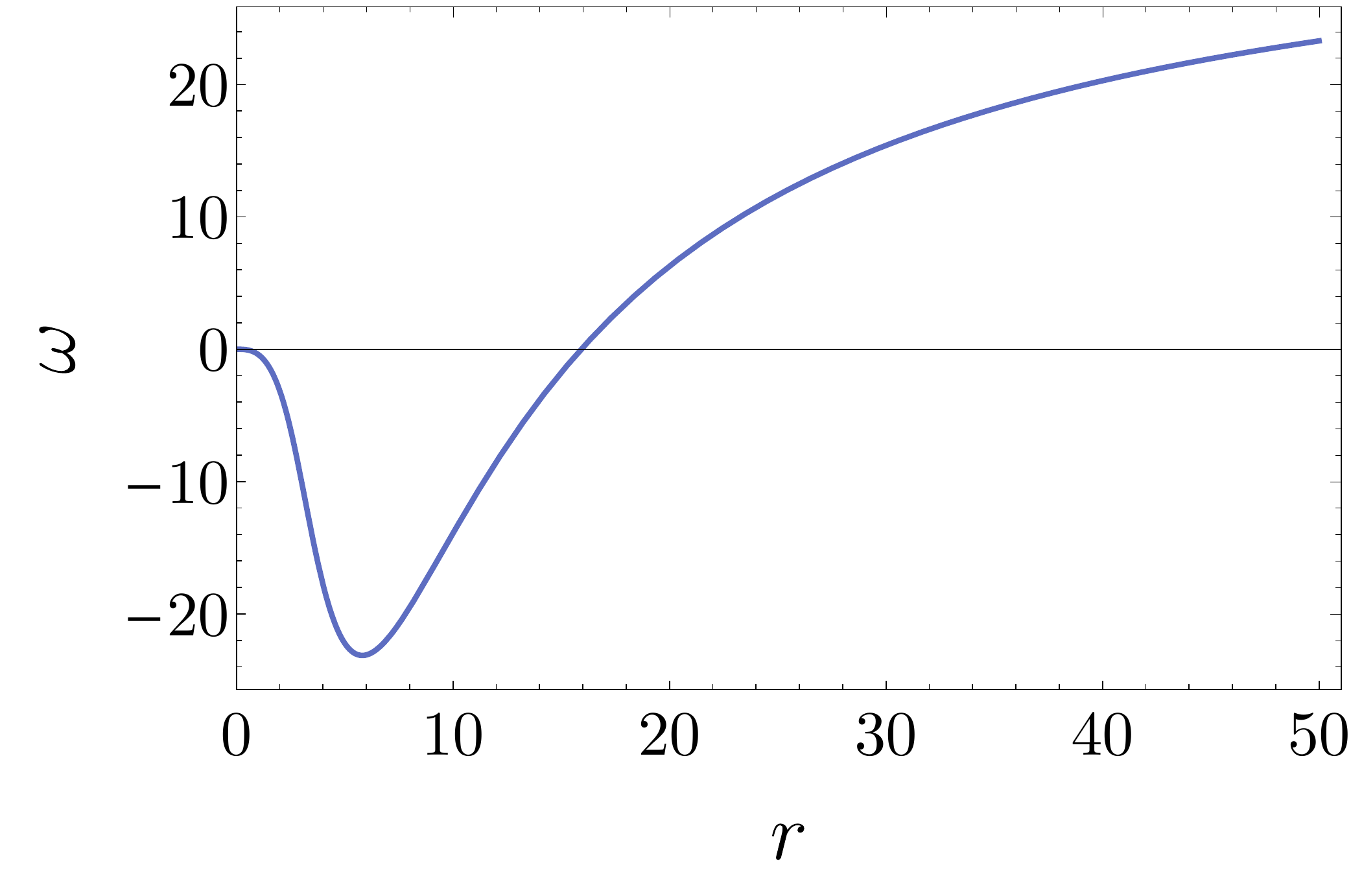}
         \caption{}
     \end{subfigure}
     \begin{subfigure}[b]{0.49\textwidth}
         \centering
         \includegraphics[width=\textwidth]{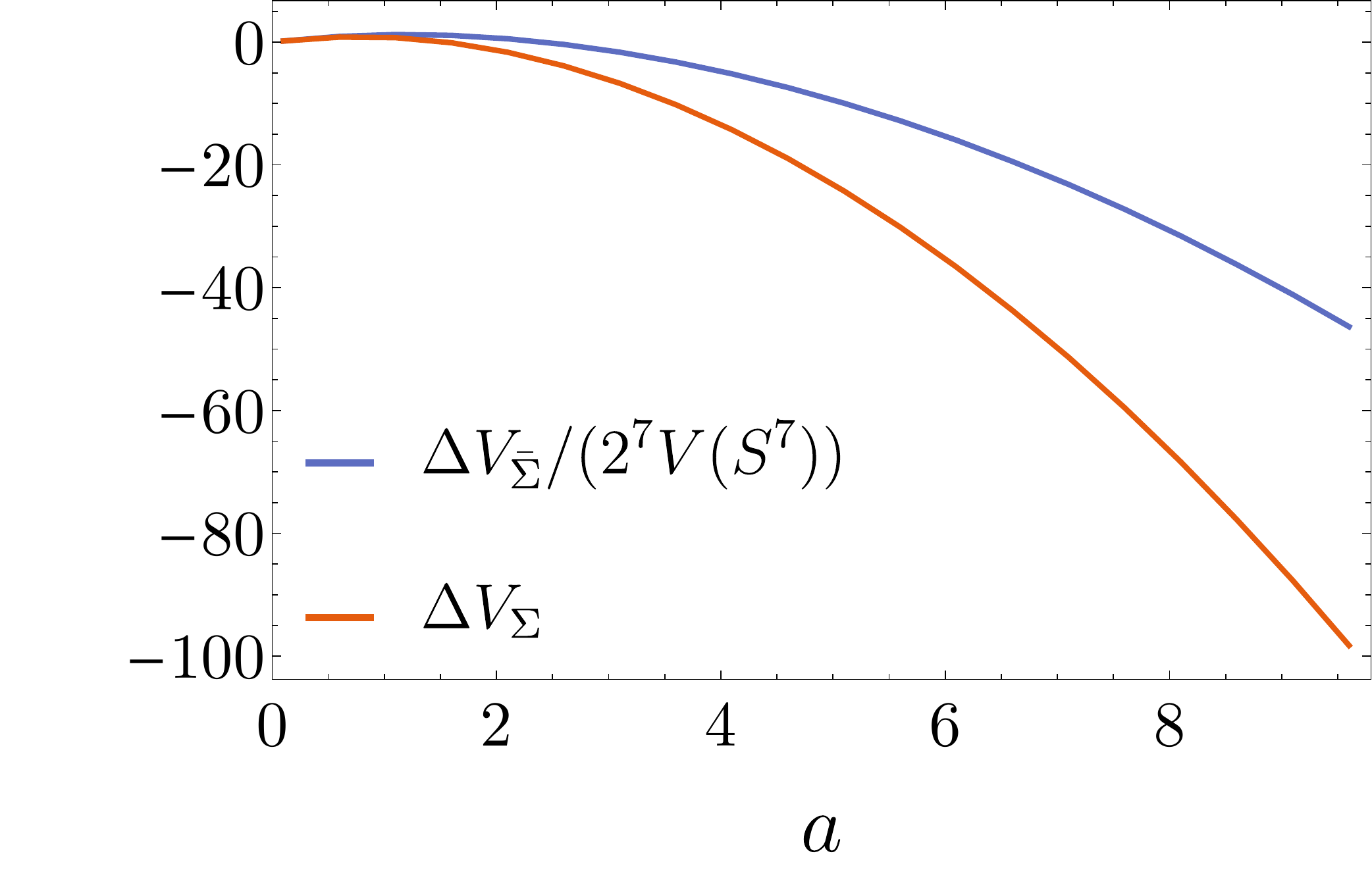}
         \caption{}
     \end{subfigure}
        \caption{(a) $\omega(r)$ for $a=5$, and (b) the vacuum-subtracted volume as function of $a$ in units of $L$.}
        \label{fig:dVplot}
\end{figure}

In Fig.~\ref{fig:dVplot} we plot the result: we show $\Delta V_{\Sigma}$ and
$\Delta V_{\tilde{\Sigma}}/V(S^7)2^7$ plotted against $a$, together with the
profile $\omega(r)$ for the value $a=5$. Other values of $a$ give a
qualitatively similar shape for $\omega(r)$.  We see that the vacuum subtracted
volume for $\Sigma$ becomes negative as we increase $a$. We find no signs of this decrease stopping for
very large values of $a$. However, if we impose a finite cutoff at $r\sim \frac{
    1 }{ \epsilon }$, the decrease will saturate at $a\sim 1/\epsilon$. Either
way, we see that turning on an increasingly large condensate of our tachyonic
scalar field  (or in the CFT language, turning on an increasingly large VEV for a relevant scalar primary) takes us closer to the reference state $\ket{R}$.
This is a very distinct behavior which a CFT dual to volume ought to reproduce.

What is the bulk mechanism leveraged in these examples that allow negative
$\mathcal{C}_F$? In the WEC-violating case ($d=3$) it is the absence of a lower
bound on the intrinsic Ricci scalar of maximal volume slices. It is illustrative
to look at a conjecture of Schoen (which is proven for $d=3$ \cite{Per02, Per03, AgoDun05}), which states
that\footnote{The conjecture is phrased in a different but equivalent way in
\cite{Schoen}. The version stated here can be found in~\cite{Bra97}.}
\begin{conj}[\cite{Schoen}]\label{conj:Schoen}
    Let $(\Sigma, h_0)$ be a closed hyperbolic Riemannian manifold with constant
    negative scalar curvature $R[h_0]$. Let $h$ be another metric on $\Sigma$
    with scalar curvature $R[h] \geq R[h_0]$. Then $\vol[\Sigma, h] \geq
    \vol[\Sigma, h_0]$.
\end{conj}
\noindent While this conjecture pertains to compact rather than conformally compact
manifolds, there are good reasons to believe it holds for conformally compact
manifolds with AdS asymptotics, as discussed in \cite{EngFol21a}. Now, in our $d=3$ data
the inequality $R[h]\geq R[h_0]$ with $h_0$ being the metric of hyperbolic space
of radius $L=1$ no longer holds as a consequence of the tachyon condensate, so this is presumably what allows 
$\mathcal{C}_F<0$. What about the eleven-dimensional case? Since the WCC allows
us to put a lower bound on the intrinsic Ricci scalar of $\tilde{\Sigma}$, it
might look like the conditions of Schoen's theorem hold. But this is not so:
the comparison manifold must have a hyperbolic metric $h_0$, and
static slices of AdS$_{4} \times S^7$ are not hyperbolic. Type IIA, IIB 
and eleven-dimensional SUGRA all satisfy (see
Appendix~\ref{sec:WECsupergravity})\footnote{Thus any AdS vacuum of these
theories will satisfy the WCC \eqref{eq:WCC} with the relevant AdS radius.}
\begin{equation}\label{eq:WEC}
\begin{aligned}
    t^a t^b \left(R_{ab}[g] - \frac{ 1 }{ 2 }g_{ab}R[g] \right) \geq 0,
\end{aligned}
\end{equation}
which through the Gauss-Codazzi equation implies that the intrinsic Ricci scalar
of any extremal hypersurface in every solution of these theories is positive.
Thus, hyperbolic volume comparison theorems no longer apply.\footnote{Strictly
speaking they could apply for some choice of $h_0$, but not when we pick $h_0$ to be a solution of the maximal-volume constraints in our theory, which is the relevant case for
$\mathcal{C}_F$.} While we do not know of any volume comparison results for
asymptotically AdS$_{d+1}\times K$ type manifolds satisfying \eqref{eq:WEC}, it is instructive
to note that volume comparison results for manifolds of spherical topology tend
to imply \textit{lower volume} when the Ricci scalar is higher \cite{Bra97}.\footnote{Note
however that in this case, a bound on just the Ricci scalar is not sufficient for
volume comparisons. Further bounds on $R_{ab}$ must be satisfied \cite{Bra97}. } This
together with our example indicates that, with respect to volume, deformations
that mainly affect the compact dimensions behave very different from those that
mainly deform the non-compact dimensions.

Finally, let us inquire about the fate of our very low $\mathcal{C}_V$ data upon
time-evolution. A guess, in keeping with earlier work on the same tachyonic
scalar theory~\cite{HerHor04} and the spacetimes considered in the next section, 
would be that it collapses into a big crunch singularity. The
negative and unbounded potential $V(\phi)\sim 1 - \cosh \sqrt{2} \phi $ seems to
favor such a collapse. However, our data is different from that of
\cite{HerHor04} and the next section in a
significant way: ours has only normalizable modes turned on, so that the
ordinary definition of the energy is finite and positive. Furthermore, the source of the decay to
a big crunch in \cite{HerHor04, HerHor05} was argued to be the presence of a lower unbounded
triple trace term in the dual theory Hamiltonian caused by the non-normalizable mode, but here this term is
not turned on due to the faster scalar field falloff. It seems likely that our data evolves to eventually form a black hole.

The skeptical reader may at this point refuse to take such unboundedly low
$\mathcal{C}_{F}$ in single-sided spacetimes seriously,  pointing out that the
CV-proposal was in its original formulation intended to describe wormholes and
the volume in the interior of a horizon. Possibly, such a reader may concede,
there are some subtleties in one-sided geometries; but surely wormholes -- the
original motivation for CV -- are still safe. 

Any such perspective is however about to be disappointed: a small modification
of the construction above allows us to build two-boundary geometries supported
by tachyonic scalars with unboundedly small vacuum-subtracted volumes.  To do
so, we modify our construction above by using the following solution to the
constraint equations
\begin{equation}
\begin{aligned}
    \omega(r) = \frac{ 1 }{ 2 }e^{-\frac{ 1 }{2} \int_{r_0}^{r} \dd \rho \rho \phi'(\rho)^2} \left\{ \omega_0 +
    \int_{r_0}^{r}\dd \rho \rho^2 e^{\frac{ 1 }{ 2 }\int_{r_0}^{\rho}\dd z
    \phi'(z)^2}\left[\left(1+ \rho^2 \right)\phi'(\rho)^2 + 2 - 2 \cosh\sqrt{2}\phi \right] \right\},
\end{aligned}
\end{equation}
where $1+r_0^2 - \frac{ \omega_0 }{ r_0 }=0$. For any $r_0$ we can use a profile similar to
\eqref{eq:scalarprofile4D} to make $\vol[\Sigma]$ arbitrarily negative compared to two copies of pure AdS. 
Thus the phenomenon of lower unbounded $\mathcal{C}_F$ is equally relevant for
wormholes.

\section{Negatively Divergent $\mathcal{C}_F$}\label{sec:vacdecay}
The clear culprit for negative $\mathcal{C}_F$ in Sec.~\ref{sec:SUGRA} was
compact dimensions or tachyonic scalars. While the main point of interest in this article is the effect of including the
compact dimensions, the importance of the scalar tachyons above clearly bears some further
investigation. In this section we provide additional
examples of tachyonic scalars causing unusual volume behavior. Previous work~\cite{Moo17b} has conducted a
near-boundary analysis that found that turning on boundary sources for these tachyons (thus changing the asymptotic
structure, in contrast with the previous section in which the asymptotics were unmodified) can result in initial data
that has divergent $\dot{\cal C}_{F}$. Our results in this section support this conclusions and
expand it further by (1) providing a full spacetime evolution to clarify the physical picture and (2) constructing
wormholes with the same properties. 

The setting will be unstable asymptotically AdS spacetimes undergoing decay. We will leverage
that there are analytical examples of such spacetimes, rather than just initial
data. The price we pay is that (1) the theory under consideration does not come
from a known realization of AdS/CFT, and (2) the scalar potential is only
known numerically. We do not expect this price to be conceptually meaningful: spacetimes that are entirely analogous qualitatively can be constructed numerically directly from the SUGRA potential \cite{HerHor04,
HerHor05}. Here we prefer to work with an analytically known metric, but we do not expect any of the qualitative features of our analysis to change by the modification of the potential. As emphasized above, we here deviate from the setup of the previous section: the scalar field
falloff, which is sufficiently slow that boundary sources are turned on, resulting in a divergent Balasubramanian-Kraus \cite{BalKra99b} stress tensor. Defining a boundary stress tensor then requires additional counterterms involving the scalar
field \cite{HenMar02, Hor03, HenMar04, HerHor04, HerMae04, HenMar06}. 
As we will see, this in turn causes the divergence structure of extremal surface volumes to differ from pure
AdS, leading to a UV-divergent $\mathcal{C}_F$ and $\frac{\dd \mathcal{C}_V}{\dd
t}$.

\subsection{AdS Vacuum Decay}
\begin{figure}
\centering
\includegraphics[width=0.35\textwidth]{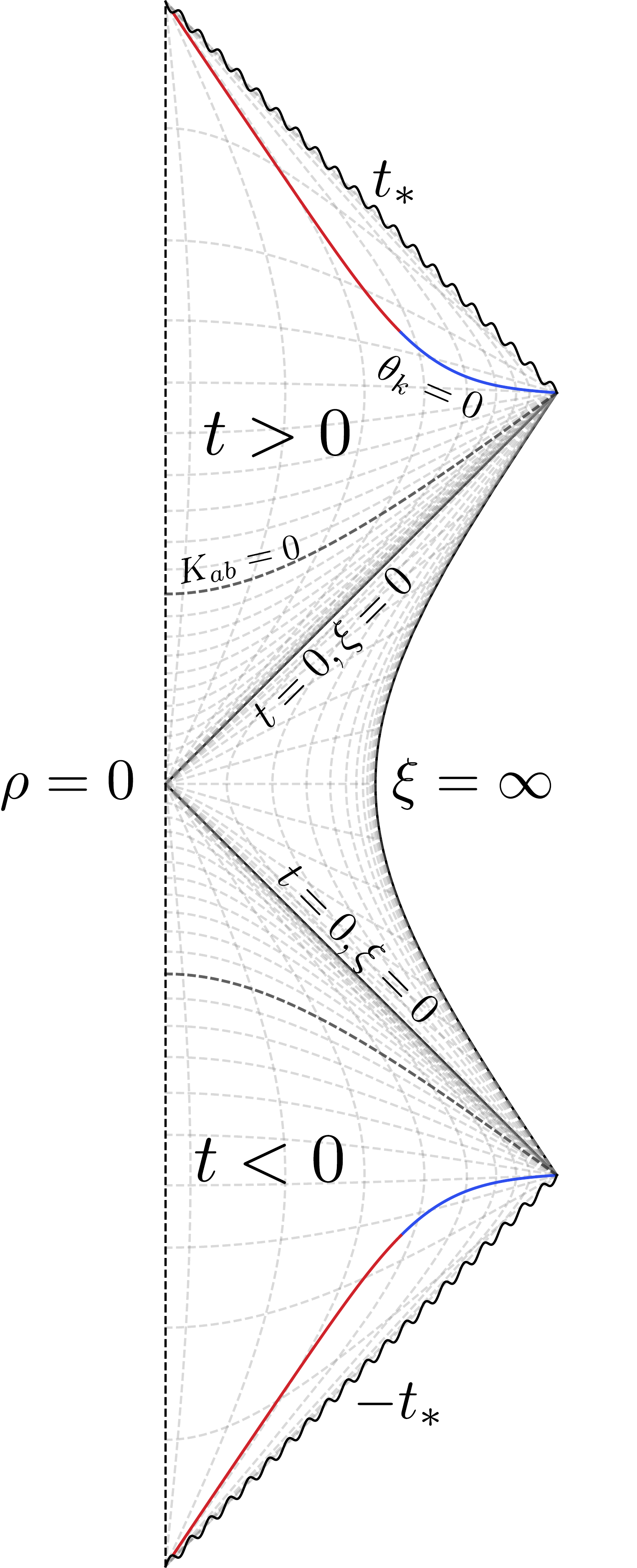}
    \caption{Conformal diagram of the spacetime \eqref{eq:CDLmetric} with $d=3$ and $c=1$. Dashed lines running vertically are hypersurfaces
    of constant $\rho$ and $\xi$, with equidistant coordinate spacing. Dashed lines running horizontally are hypersurfaces
    of constant $t$ and $\zeta$, with equidistant coordinate spacing.  The blue line is the spacelike section of a
    holographic screen, while the red line is the timelike portion. The darker horizontal dashed lines are the constant$-t$
    surfaces at which the FRW region transitions between expanding and crunching ($a'(t)=0$), which are totally
    geodesic. These are extremal surface barriers.
    }
\label{fig:conformaldiagram}
\end{figure}
The one-sided spacetimes we consider are given by the one-parameter family of
metrics constructed in \cite{DonHar11}, parametrized by the real positive
parameter $c$. These geometries are covered by two coordinate patches, with patch I having metric 
\begin{equation}\label{eq:CDLmetric}
\begin{aligned}
    \dd s^2_{\rm I} &= \dd \xi^2 + a(\xi)^2\left(-\dd \zeta ^2 + \cosh^2 \zeta \dd \Omega^2 \right),\\
%    \dd s^2_{\rm II} &= - \dd t^2 + \hat{a}(t)^2 \left( \dd \rho^2 +  \sinh^2 \rho \dd \Omega^2\right), \\
\end{aligned}
\end{equation}
where $\dd \Omega^2$ now is the metric of a $d-1$--dimensional sphere, and with
\begin{equation}\label{eq:afuncs}
\begin{aligned}
    a(\xi) &= (1+c)\sinh \xi - 2c \sinh \frac{ \xi }{ 2  }. %\\
    %\hat{a}(t) &= (1+c) \sin t - 2 c \sin \frac{ t }{ 2 }. \\
\end{aligned}
\end{equation}
The second patch, with coordinates $(t, \rho, \Omega_i)$, is obtained by the analytic continuation $\xi=i t, \tau
= i\rho$.
Patch I is the causal wedge of the spacetime, with the AdS
boundary at $\xi=\infty$, where the scalar field is in the false vacuum (a local
maximum). As $\xi$ approaches 0, where the edge of the causal wedge lies, the scalar field approaches the true
vacuum value.  Patch II is an FRW-region which initially expands with motion away from $t=0$ and then crunches,
with a curvature singularity at $t=\pm t_{*}$,
where $a(i t_{*})=0$. Physically, the $\zeta=0$ slice is a bubble of true vacuum that nucleates inside the false vacuum at a moment of
time symmetry.  Forward or backward time-evolution results in the subsequent decay of
the false AdS vacuum. See Fig.~\ref{fig:conformaldiagram} for a conformal
diagram drawn for the case $c=1, d=3$ (see Appendix~\ref{sec:CDLapp}
for the computation). 

This spacetime has
past and future cosmological singularities, and in the boundary
conformal frame of the static cylinder, the boundary exists only for a finite
time. The dual field theory, if it exists, can either be viewed as living on de
Sitter space, or it can be seen as a field theory on the Einstein static
universe whose evolution terminates in finite time -- possibly due to a
Hamiltonian that is unbounded below, as discussed in~\cite{HerHor04}.

\subsubsection*{$\mathcal{C}_F$ is negative and divergent}
Consider again the coordinates
\begin{equation}\label{eq:cancoord}
\begin{aligned}
    \dd s^2|_{\Sigma} = \frac{ 1 }{ 1 + r^2 - \frac{ \omega(r) }{ r^{d-2} }  }\dd r + r^2 \dd \Omega^2,
\end{aligned}
\end{equation}
on $\Sigma$.
Assume now that $\omega(r)$ is divergent at large $r$, with the leading behavior at large$-r$
given by $\omega(r)\sim \omega_s r^s$ for $0<s<d$.\footnote{$s\geq d$ is
incompatible with being asymptotically AdS with radius $1$.} The leading $\omega$-dependent divergence in the
volume then is given by:
\begin{equation}
\begin{aligned}
    \vol[\Sigma] &\sim \vol[S^{d-1}] \int^{r_{\rm cut}} \dd r \frac{ r^{d-1} }{ \sqrt{1 + r^2 - \frac{ \omega(r) }{
    r^{d-2}}} } \\
    &\sim \vol[S^{d-1}] \int^{r_{\rm cut}} \dd r r^{d-2} \left[\frac{ 1 }{ 2 r^{d-s} }\omega_{s} + \ldots \right] \\
    &\sim \omega_{s} \frac{ \vol[S^{d-1}] r^{s-1}_{\rm cut} }{ 2(s-1) },
\end{aligned}
\end{equation}
Comparing with a slice of pure AdS with cutoff at the same area-radius $r_{\rm
cut}$, we find
\begin{equation}
\begin{aligned}
    \vol[\Sigma] - \vol[\Sigma_{\text{AdS}_{4}}] = \omega_{s} \frac{
    \vol[S^{d-1}] r^{s-1}_{\rm cut} }{ 2(s-1) } + \text{subleading}.
\end{aligned}
\end{equation}
We now proceed to show that for our spacetime, we have $s=d-\frac{ 1 }{ 2 }$ and
$\omega_s < 0$, giving that $\mathcal{C}_F$ is negative and divergent. 

To calculate $\omega_s$, it is useful to know that there is a geometric
functional $\omega[\sigma, \Sigma]$ that reduces to $\omega(r)$ when $\sigma$
is a symmetric codimension$-2$ spatial surface and $\Sigma$ spherically
symmetric \cite{EngFol21a}:\footnote{This is proportional to the so-called Geroch-Hawking mass when
$d=3$.}
\begin{equation}
\begin{aligned}
    \omega[\sigma, \Sigma] &= \frac{ 1 }{ \vol[S^{d-1}] }
     \left(\frac{ A[\sigma] }{\vol[S^{d-1}] }\right)^{\frac{ 1 }{ d-1
             }}\int_{\sigma}\left[\frac{ \mathcal{R} }{ (d-1)(d-2) } - \frac{ H^2 }{ (d-1)^2 } + \frac{ 1
    }{L^2 }\right],
\end{aligned}
\end{equation}
where $H[\sigma]$ is the mean curvature of $\sigma$ inside $\Sigma$ and
$\mathcal{R}$ the intrinsic Ricci scalar of $\sigma$. Let now $\Sigma$ be the
$\zeta=0$ hypersurface, which is a maximal volume slice since it is a
moment of time symmetry. Evaluating $\omega[\sigma, \Sigma]$ for a constant
$\xi$ surface $\sigma$, we find
\begin{equation}
\begin{aligned}
    \omega(\xi)= a(\xi)^{d - 2}\left(1 + a(\xi)^2 - a'(\xi)^2\right) 
    =  - \frac{ 1 }{ 2^{d-1} }c(1+c)^2 e^{\left(d- \frac{ 1
     }{ 2 } \right)\xi }  + \mathcal{O}(e^{(d-1)\xi}).
\end{aligned}
\end{equation}
If we were to change coordinates to the form \eqref{eq:cancoord} we would find
$r = \mathcal{O}(e^{\xi})$, and so indeed we have $s=d-1/2$, $\omega_{s}<0$,
showing that $\mathcal{C}_F$ is negative and divergent.

Note that pure AdS-subtracted volume here is somewhat unnatural
from the field theory perspective. The scalar field falls off sufficiently slowly so
as to turn on a source on the boundary: pure AdS is not 
a solution of the boundary theory dual to \eqref{eq:CDLmetric}, so the comparison appears ill-motivated. In this particular setting -- though not in the previous section -- the
negatively divergent $\mathcal{C}_F$ should be viewed as a
statement purely about volumes in asymptotically AdS spacetimes, rather than as a statement
pertaining to a single field theory. It is in principle possible that the field
theory dual to \eqref{eq:CDLmetric}, if it exists, has a preferred state for the volume subtraction, for which $\mathcal{C}_F$ would be positive. 

\subsubsection*{Complexity change}
Let us now compute the leading divergent contribution to the complexity change
for spherically symmetric maximal volume slices.
To compute the change in complexity we must choose the bulk cutoff carefully. Any given conformal frame induces a unique Fefferman-Graham coordinate
system in a neighbourhood of the conformal boundary \cite{FefGra85,GraLee91,GraWit99}:
\begin{equation}
\begin{aligned} 
    \dd s^2 = \frac{ 1 }{ z^2 }[\dd z^2 + \gamma_{\mu\nu}(z, x)\dd x^{\mu}\dd
    x^{\nu}],
\end{aligned}
\end{equation}
where $z=0$ is the conformal boundary and $\gamma_{\mu\nu}(0, x)$ the chosen
conformal representative. Given this coordinate system, we can cut off volumes at $z=\epsilon$. 

In the case at hand there are two natural conformal frames; we can either choose
the boundary to be dS$_d$ or the static cylinder. With respect to the dS$_d$ conformal frame it 
can readily be checked that the leading divergence in the volume of the maximal volume slice anchored at a
constant $\zeta$ is
\begin{equation}
\begin{aligned}
    \vol[\Sigma_{\zeta}] = \frac{ \vol[S^{d-1}]\cosh \zeta^{d-1} }{(d-1)
        \epsilon^{d-1} }+\mathcal{O}\left(\epsilon^{d-3/2}\right).
\end{aligned}
\end{equation}
This increases to the future and past of $\zeta=0$ simply because 
(1) when regulating with a Fefferman-Graham cutoff, the divergence of maximal volume slices is always proportional to
the boundary volume in the chosen conformal frame, and (2) the volume of constant $\zeta$ slices of de Sitter
increases to the future and past.

Next, let us look at the more interesting case of the static cylinder conformal frame. 
A computation (see appendix~\ref{sec:CDLapp}) of the leading order complexity change with cutoff adapted to the static
cylinder gives
\begin{equation}\label{eq:staticCyl}
\begin{aligned}
    \frac{ \dd \vol[\Sigma_t] }{ \dd t } =  -\frac{ 1 }{ \epsilon^{d-\frac{ 3 }{ 2 }} } \frac{ d-1 }{ d-\frac{ 3 }{ 2 } }
    \sqrt{\frac{ c^2 }{ 2(c+1) }} \frac{ \sin t }{ (\cos t)^{3/2}  } +
    \mathcal{O}(\epsilon^{-d+2}) , \qquad t\in \left(-\frac{ \pi }{ 2 }, \frac{ \pi }{
    2}\right).
\end{aligned}
\end{equation}
This is clearly negative and divergent, and so unlike in well known examples of
black holes, the moment of time-symmetry is here a maximum of $\mathcal{C}_V$, rather than a minimum.
This is in constrast with the case of the dS conformal frame, and so
highlights how extremal hypersurface volume is an observable whose  UV-divergence structure depends strongly on the choice of
boundary conformal frame. This is consistent with \cite{Moo17b}'s near-boundary analysis, which also found a divergent rate of change for tachyonic
scalars with boundary sources turned on. In our example we have the
additional benefit of knowing the spacetime globally, providing a physical
picture of what is happening in the bulk. 

The decrease of $\mathcal{C}_V$ is a UV effect, and the volume behind the
horizon is admittedly increasing towards the future (before the crunch region
that is, which is anyway hidden from all extremal surfaces). However, the volume behind the horizon is not
a natural observable to associate to a boundary state at a fixed time,
since turning on sources in the future would alter the horizon location, and
thus also the volume behind it. Nevertheless, it does seem that the CV 
proposal needs some modification in the spacetimes considered in this section.
One possibility is some generalized volume functional that includes contributions
from the scalar fields, which from a dimensional reduction perspective this appears
natural. 

\subsection{Decaying Cosmological Wormholes}
\begin{figure}[htb]
\centering
\includegraphics[width=0.6\textwidth]{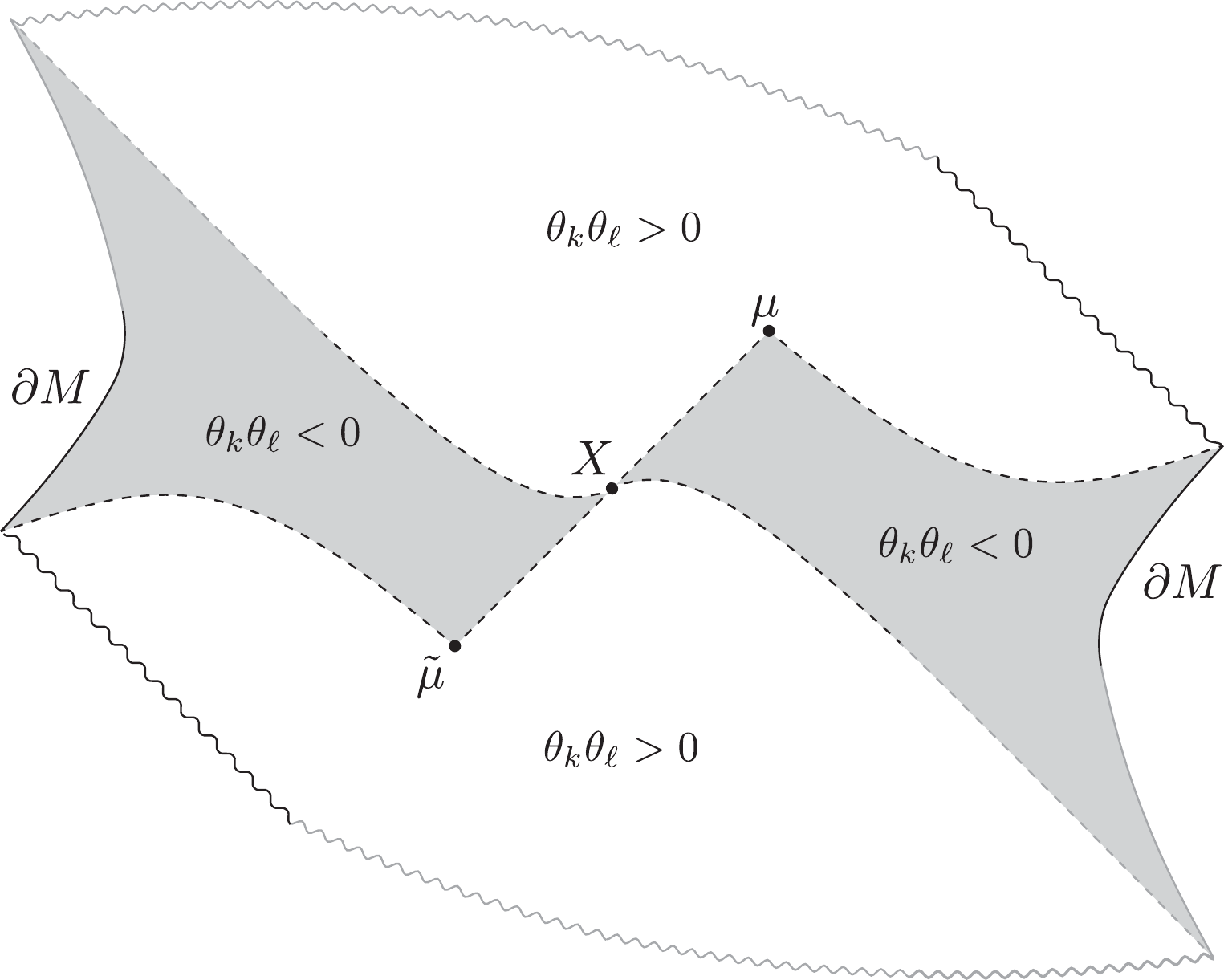}
    \caption{Black lines show numerically computed features of a coarse-grained 
    spacetime formed from a marginally trapped (minimar \cite{EngWal18}) surface $\mu$ on the spacelike section of the
    holographic screen shown Fig.~\ref{fig:conformaldiagram}. Dashed lines represent
    apparent horizons, while $X$ is the HRT surface. Grey lines lie outside the range of our numerics, and are pure sketches representing a 
    qualitative image of how the full spacetime could look like. For an illustration with additional details, see
    Fig.~\ref{fig:cosmowormhole} in the
    appendix. }
\label{fig:cosmowormhole_penrose}
\end{figure}

Can we find a wormhole with the same properties as the spacetime considered above -- a wormhole with cosmological
singularities, negative divergent complexity of formation, and decreasing late
time complexity? 
That would appear to be in some tension with the paradigm of wormhole volume corresponding to increasing complexity; such an example would add urgency in finding an appropriate modification of CV that can accommodate such spacetimes. 

It turns out that we can, in fact, build such a spacetime. The procedure is
borrowed from  \cite{EngWal17b, EngWal18}, which constructs a two-boundary
wormhole from a single-sided spacetime containing a marginally trapped surface
satisfying certain assumptions. The protocol is roughly as follows: one fixes
the data in the exterior of a given marginally trapped surface $\mu$; the rest
of the initial data is provided on a stationary null hypersurface fired
towards the past interior from $\mu$. This hypersurface eventually develops an extremal surface on it; at this surface, the initial data is CPT conjugated, resulting in a second boundary. The initial data is characteristic; in our case, fixing a spherically symmetric marginally trapped surface means that obtaining the spacetime requires numerical evolution of the spherically symmetric characteristic 
Einstein-Klein-Gordon equations.  We will not expound on the details of the numerics here (although they are surprisingly simple),
instead just briefly summarizing our results. Figure~\ref{fig:cosmowormhole_penrose} illustrates one of these wormholes. 
On the left, we show the marginally trapped surface $\mu$ used to build the wormhole embedded in the original spacetime.
On the left we see the coarse-grained spacetime in the regions where we are able to obtain it numerically. 
We have not obtained any parts of the spacetime in the future or past of $X$, since this requires evolution past a
shockwave, requiring more sophisticated methods than our fairly straightforward technique. The upshot of this solution
is that spacetime emergence connecting two asymptotic boundaries via an interior need \textit{not} feature a simple
complexity growth with time, emphasizing the necessity of a refinement to the Complexity=Volume proposal  that takes
into account the different behaviors of deformed theories. We emphasize that this is
qualitatively (as well as quantitatively) different from the negative ${\cal C}_{F}$ of the previous section, in which no boundary sources were turned on.

\section*{Acknowledgments}
It is a pleasure to thank Jan de Boer, Shira Chapman, Sebastian Fischetti, Patrick Jefferson, Lampros Lamprou, Hong Liu, Dan Roberts, and
Wati Taylor for discussions.  This work is supported in part by NSF grant no.
PHY-2011905 and the MIT department of physics. The work of NE was also supported in part by the U.S. Department
of Energy, Office of Science, Office of High Energy Physics of U.S. Department of Energy under grant Contract Number
DE-SC0012567 (High Energy Theory research) and by the U.S. Department of Energy Early Career Award DE-SC0021886. The work of \AA{}F is also
supported in part by an Aker Scholarship. 

\appendix
\section{Appendix}\label{sec:appendix}
\subsection{The DEC and SEC in type II and $D=11$ SUGRA}\label{sec:WECsupergravity}

\subsubsection*{The DEC and the SEC for $p$-forms}
Consider a stress tensor
\begin{equation}\label{eq:pformstress}
\begin{aligned}
    T_{ab} = \frac{ 1 }{ p! }\left[p F\indices{_{a}^{c_2 \ldots c_p}}F_{b c_2 \ldots c_p} 
    - \frac{ k }{ 2 }g_{ab}F^{c_1 \ldots c_{p}}F_{c_1 \ldots c_{p}} \right], \qquad 1 \leq k\leq p,
\end{aligned}
\end{equation}
where $F_{c_1 \ldots c_p}$ is a $p$-form. For $k=1$ this is just the stress tensor of a free $(p-1)$ form with curvature
$F_{c_1 \ldots c_{p}}$ and action
$S=-\frac{ 1 }{ 2 }\int F \wedge \star F$. We first check the DEC. Let $u^a, v^b$ be any two timelike vectors at
a point $p$. We will normalize them to our convenience, since the DEC is independent of the choice of
normalization. Choose Riemann normal coordinates at $q$, so that $g_{\mu\nu}|_{q} = \eta_{\mu\nu}$ and
$u^{\mu}=(\partial_{t})^{\mu}$, where $\eta_{\mu\nu}$ is the Minkowski metric in Cartesian coordinates. Next, we can always rescale $v$ and perform a rotation of our coordinates so that
\begin{equation}
\begin{aligned}
    v^{\mu} = (\partial_t)^{\mu} + f(\partial_x)^{\mu},
\end{aligned}
\end{equation}
for some constant $f\geq0$. With this, we now find at $q$ that
\begin{equation}
\begin{aligned}
v^{a} u^{b} T_{ab} = T_{tt} + f T_{tx}.
\end{aligned}
\end{equation}
We compute that
\begin{equation}\label{eq:Ttt}
\begin{aligned}
    p! T_{tt} &=  \sum_{\mu_i}\left( pF\indices{_{t}^{\mu_2 \ldots \mu_{p}}}F_{t\mu_2 \ldots \mu_{p}} + \frac{ k }{ 2 }F^{\mu_1 \ldots \mu_p}F_{\mu_1
    \ldots \mu_{p}}\right) \\
    &= \left(p - \frac{ k }{ 2 }\right)\sum_{\mu_i} F_{t \mu_2 \ldots \mu_p}^2
    + \frac{ k }{ 2 }\sum_{\mu_i, \mu_1 \neq t} F_{\mu_1 \ldots \mu_p}^2 \\
    &\geq \left(p - \frac{ k }{ 2 }\right)\sum_{\mu_i} F_{t\mu_2 \ldots \mu_p}^2
    + \frac{ k }{ 2 }\sum_{\mu_i} F_{x\mu_2 \ldots \mu_p}^2,
    %+ \frac{ 1 }{ 2 }\sum_{\mu_1 \neq \{t, x\}} F_{\mu_1 \ldots \mu_p}^2 \\
\end{aligned}
\end{equation}
and 
\begin{equation}\label{eq:Ttx}
\begin{aligned}
    p! T_{tx} &= \sum_{\mu_i }p F\indices{_{t}^{\mu_2 \ldots \mu_p}} F_{x \mu_2 \ldots \mu_p}
    &= p \sum_{\mu_i} F_{t\mu_2 \ldots \mu_p}F_{x \mu_2 \ldots \mu_p}.
\end{aligned}
\end{equation}
And so adding up \eqref{eq:Ttt} and \eqref{eq:Ttx} we get 
\begin{equation}
\begin{aligned}
    p! T_{ab}u^a v^b &\geq p \sum_{\mu_i}
    \left[\left(1-\frac{ k }{ 2p } \right)F_{t\mu_2 \ldots \mu_p}^2 
    + \frac{ 1 }{ 2 }f F_{t\mu_2 \ldots \mu_p}F_{x\mu_2 \ldots \mu_p} + \frac{ k }{ 2 }F_{x\mu_2 \ldots \mu_p}^2
    \right] \\
    &\qquad\qquad + \frac{ 1 }{ 2 }pf \sum_{\mu_i }F_{t\mu_2 \ldots \mu_p}F_{x \mu_2 \ldots \mu_p} \\
    & \geq p \sum_{\mu_i }\left[\left(1-\frac{ k }{ 2p } \right)F_{t\mu_2 \ldots \mu_p}^2 
    + \frac{ 1 }{ 2 }f F_{t\mu_2 \ldots \mu_p}F_{x\mu_2 \ldots \mu_p} + \frac{ k }{ 2 }F_{x\mu_2 \ldots \mu_p}^2
    \right].
\end{aligned}
\end{equation}
This is non-negative for each term in the sum. If $F_{t\mu_2 \ldots \mu_p}F_{x \mu_2 \ldots \mu_p}\geq 0$ this is obvious,
since we assumed $1 \leq k\leq p$, giving that each term is manifestly non-negative. 
So assume $F_{t\mu_2 \ldots \mu_p}F_{x \mu_2 \ldots \mu_p}<0$. In this case we get a 
smaller term if we replace (1) $f\rightarrow 2$, (2) $k \rightarrow p$ in the first term, and (3) $k\rightarrow 1$ in the last term:
\begin{equation}
\begin{aligned}
    p! T_{ab}u^a v^b &
    \geq p \sum_{\mu_i }\left[ \frac{ 1 }{ 2 }F_{t\mu_2 \ldots \mu_p}^2 
    + F_{t\mu_2 \ldots \mu_p}F_{x\mu_2 \ldots \mu_p} + \frac{ 1 }{ 2 }F_{x\mu_2 \ldots \mu_p}^2
    \right] \geq 0.
\end{aligned}
\end{equation}
Thus the DEC holds for the stress tensor \eqref{eq:pformstress}.
Together with $G_{ab}=8\pi G_N T_{ab}$, this implies the WCC \eqref{eq:WCC}.

Next, let us turn to the SEC. Set $8\pi G_N=1$. Then
\begin{equation}
\begin{aligned}
   T \equiv  T\indices{^{a}_{a}} = g^{ab}G_{ab} = \left(1 - D/2\right)R
\end{aligned}
\end{equation}
and so
\begin{equation}
\begin{aligned}
    R_{ab}t^{a}t^{b} &= \left(G_{ab} +\frac{ 1 }{ 2 }g_{ab}R\right)t^{a}t^{b}
    &= T_{ab}t^{a}t^{b} - \frac{ 1 }{ 2 }R 
    &= T_{ab}t^{a}t^{b} - \frac{ 1 }{ 2-D }T.
\end{aligned}
\end{equation}
Our stress tensor gives
\begin{equation}
\begin{aligned}
    p! T &= \frac{ 1 }{ 2 }\left( 2p - kD\right) F^{c_1 \ldots c_p}F_{c_1 \ldots c_p}.
\end{aligned}
\end{equation}
Choosing again Riemann normal coordinates at our point of interest, we get
\begin{equation*}
\begin{aligned}
    p! R_{\mu\nu}t^{\mu}t^{\nu} &= \left(p - \frac{ k }{ 2 }\right)\sum_{\mu_i} F_{t\mu_2 \ldots \mu_p}^2 + \frac{ k }{
        2 }\sum_{\mu_i, \mu_1 \neq t}F_{\mu_1
    \ldots \mu_p}^2 -\frac{ 1 }{ 2 }\frac{ 2p-k D }{ 2-D }F_{\mu_1 \ldots \mu_p}F^{\mu_1 \ldots \mu_p} \\
    &\geq \left(p - \frac{ k }{ 2 } - \frac{ 1 }{ 2 } \frac{ k D - 2p  }{ D -2 }\right)\sum_{\mu_i} F_{t\mu_2 \ldots
    \mu_p}^2 \\
    &\geq \left(p - \frac{ p }{ 2 } - \frac{ 1 }{ 2 } \frac{ p D - 2p  }{ D-2 }\right)\sum_{\mu_i} F_{t\mu_2 \ldots
    \mu_p}^2 \\
    & = 0,
\end{aligned}
\end{equation*}
where we used that $1 \leq k\leq p$ above.
Thus the SEC holds for the stress tensor \eqref{eq:pformstress}.

\subsubsection*{Type IIB}
From~\cite{LuPop99, CasDal10} we have that the gravitational equations of motion of the bosonic sector
of type IIB supergravity in the Einstein frame can be written as
\begin{equation}
\begin{aligned}
    R_{ab} =& \frac{ 1 }{ 2 }\nabla_{a}\phi \nabla_{b}\phi + \frac{ 1 }{ 2 }e^{2\phi}\nabla_a \chi \nabla_b \chi +  \frac{ 1 }{ 96 }H_{acdef}
    H\indices{_{b}^{cdef}} + \frac{ 1 }{ 4 }e^{-\phi}\left[F_{acd}F\indices{_{b}^{cd}} - \frac{ 1 }{ 12
    }g_{ab}F_{cde}F^{cde} \right] \\ 
    &+ \frac{ 1 }{ 4 }e^{-\phi}\left[L_{acd}L\indices{_{b}^{cd}} - \frac{ 1 }{ 12
    }g_{ab}L_{cde}L^{cde} \right]
\end{aligned}
\end{equation}
where $H$ is a five form and $F$ and $L$ are three-forms. The scalar stress tensors are (up to a positive rescaling in
the case of $\chi$) just stress tensors of massless scalars and so satisfies the DEC and the SEC. Thus we just need to check
that the $p$-forms. Rewriting the equation in Einstein form we get stress tensors
\begin{equation}
\begin{aligned}
    T^{(H)}_{ab} &= \frac{ 1 }{ 5\times 96 }\left( 5 H_{acdef} H\indices{_{b}^{cdef}} - \frac{ 5 }{ 2 }g_{ab}H_{cdefg}
    H\indices{^{cdefg}} \right) \\
    T^{(F)}_{ab} &= 
    \frac{ 1 }{ 12 }e^{-\phi}\left(3 F_{acd}F\indices{_b^{cd}} - \frac{ 1 }{ 2 }g_{ab}F_{cde}F^{cde}\right)
    \\
    T^{(L)}_{ab} &= \frac{ 1 }{ 12 }e^{\phi}\left(3 L_{acd}L\indices{_b^{cd}} - \frac{ 1 }{ 2 }g_{ab}L_{cde}L^{cde}\right) \\
\end{aligned}
\end{equation}
All of these stress tensors are proportional to \eqref{eq:pformstress} with a positive coefficient, and so the DEC and SEC holds.

\subsubsection*{Type IIA and $D=11$ supergravity}
In the Einstein frame, the stress tensors of the bosonic matter in type IIA and eleven-dimensional supergravity is just that of free $p$-form
fields, except for an overall positive factor proportional to an exponential of the dilaton in the case of type IIA \cite{Ham16}.
The exact kind of computation as was carried out above shows that the WEC holds classically in these theories.

\subsection{General properties of $K=0$ slices in SUGRA}\label{sec:generalProps}
We here derive and highlight some general results on
slices of vanishing mean curvature in type IIA/B and $D=11$ SUGRA. This will justify our statement that the
initial data in Sec.~\ref{sec:SUGRA} gives the unique maximal volume slice in the evolved spacetime.

The first observation is the following:
\begin{prop}\label{prop:SEC}
    The bosonic matter fields of type IIA, IIB and $D=11$ supergravity in the Einstein frame satisfies the strong energy condition (SEC) and the
    dominant energy condition (DEC):
    \begin{equation}
    \begin{aligned}
        \text{SEC}&:\qquad & T_{ab}u^a u^b + \frac{ 1 }{ D-2 }T\indices{^a_a} &\geq 0, \qquad \forall \text{timelike }
        u^a, \\
        \text{DEC}&:\qquad & T_{ab} u^a v^a &\geq 0, \qquad \forall \text{timelike } u^a, v^a.
    \end{aligned}
    \end{equation}
\end{prop}
\noindent This result is known in the literature \cite{MalNun01, Gib03,TowWoh}, but for convenience we included a
derivation above, as the result is often stated without proof. Note that this result applies to the standard bosonic fields,
and does not include stringy curvature corrections or additional massive fields. Also, dimensional reduction will in
general both break the SEC \cite{TowWoh} and the DEC \cite{HerHor03} (although for specific types of compactifications
they might survive \cite{Gib03}).

The well known fact that the SEC combined with $K=0$ implies maximality \cite{BriFla76, MarTip80} now immediately gives
\begin{prop}\label{prop2}
    Let $(M, g)$ be any classical solution of type IIA, IIB, or $D=11$ SUGRA in the Einstein frame. If $\Sigma$ is a $K=0$ spacelike hypersurface, possibly with boundary, then $\Sigma$ is maximal under any variation that leaves
    its boundary fixed. 
\end{prop}
\noindent This is a manifestation of the well known fact that the SEC ensures focusing of timelike congruences.
The result follows from calculating the second variation of the volume of a spacelike $K=0$ hypersurface, which reads \cite{BriFla76, MarTip80}
\begin{equation}\label{eq:secondVar}
\begin{aligned}
    \delta^2 \vol[\Sigma] &=  - \int_{\Sigma}(|DN|^2 + N^2 K_{ab}K^{ab} + N^2 R_{ab}n^a n^a) \leq 0,
\end{aligned}
\end{equation}
where $\tau^a$ is the vector field generating the variation of $\Sigma$, $n^a$ a unit normal to $\Sigma$,
$D_a$ the connection on $\Sigma$, and $N=\tau \cdot n$. It is assumed that the boundary of $\Sigma$ is kept 
fixed in \eqref{eq:secondVar}.  

As described above, Proposition \ref{prop2} will generally not be true in dimensionally reduced spacetime, and so in this case we actually have better control in the full $D=10, 11$ spacetime. 
This shows a situation in AdS/CFT where the compact dimensions should be viewed as a resource rather than a nuissance. 

Next, \cite{CouEcc18} has showed that if the SEC holds, then there cannot be two $K=0$ slices anchored at the same
boundary time in an asymptotically AdS spacetime (the proof survives when we also have a compact space). Thus we have the result
\begin{prop}\label{prop3}
    Let $(M, g)$ be an asymptotically AdS$_{d+1}\times X$ solution of type IIA, IIB, or $D=11$ SUGRA in the Einstein frame for
    some compact manifold $X$.
    Let $\Sigma$ be a complete maximal volume slice anchored at boundary time $C$. Then there is no other maximal volume
    slice in the domain of dependence of $\Sigma$ that is anchored at $C$.
\end{prop}
\noindent Proposition \ref{prop2} and \ref{prop3} now justifies our assertion from Sec.~\ref{sec:SUGRA} that our initial data $\Sigma$
is the true maximal volume slice.

Finally, we remark that Proposition \ref{prop2} and \ref{prop3} remain true if we add additional SEC-respecting
matter, such $p$-dimensional branes $B$ with action
\begin{equation}
\begin{aligned}
    S = - \int_B \dd^{p}x \sqrt{-h} T + \int_{B}C_p,
\end{aligned}
\end{equation}
where $T$ is a non-negative potentially field-dependent scalar and $C_p$ a $p$-form -- both independent of the induced
metric $h_{ab}$ on $B$.

\subsection{AdS Vacuum Decay Computations}
\label{sec:CDLapp}
\subsubsection*{Kruskal coordinates}
Define
\begin{equation}
\begin{aligned}
    \hat{a}(t) = a(i t) = (1+c) \sin t - 2 c \sin \frac{ t }{ 2 }.
\end{aligned}
\end{equation}
Consider the metric \eqref{eq:CDLmetric} in the special case of $c=1$ and $d=3$. Define the functions
\begin{equation}
\begin{aligned}
    R(\xi) &= \int_{\xi_0}^{\xi} \frac{ \dd \xi' }{ a(\xi') } = \frac{ 1 }{ 3 }\log \left[\frac{ f(\xi) }{ f(\xi_0) }\right],
    \qquad 
    T(t) = \int_{t_0}^t \frac{ \dd t' }{ \hat{a}(t') }= \frac{ 1 }{ 3 }
    \log \left[\frac{ \hat{f}(t) }{ \hat{f}(t_0) }\right] \\
\end{aligned}
\end{equation}
where 
\begin{equation}
\begin{aligned}
    f(\xi) &=  \frac{  \cosh\left( \frac{ \xi }{ 4 } \right)  \sinh \left(\frac{ \xi }{ 4 } \right)^3 
         }{ \left( 1-2 \cosh\left(\frac{ \xi }{ 2 }\right)\right)^2 
         } = \begin{cases}
             \frac{ \xi^3 }{ 64 } + \mathcal{O}(\xi^5) & \xi \ll 1 \\
             \frac{ 1 }{ 16 } - \frac{ 3 }{ 16 }e^{-\xi}- \frac{ 1 }{ 8 }e^{-3\xi/2}+\mathcal{O}(e^{-2\xi}) & \xi \gg 1
         \end{cases}\\
    \hat{f}(t) &=  \frac{  \cos\left( \frac{ t }{ 4 } \right)  \sin \left(\frac{ t }{ 4 } \right)^3 
         }{ \left( 1-2 \cos\left(\frac{ t }{ 2 }\right)\right)^2 
         } = \begin{cases}
             \frac{ t^3 }{ 64 }+ \mathcal{O}(t^5) & t \ll 1 \\
             \frac{ 1 }{ 4\sqrt{3}(t_*-t)^2 } - \frac{ 1 }{ 8(t_* - t) } + \mathcal{O}(1) & t_* - t \ll 1 
         \end{cases}
\end{aligned}
\end{equation}
where $t_* = \frac{ 2\pi }{ 3 }$ is the location of the future singularity, ie.\ the maximal value of $t$. The constants
$t_0>0$ and $\xi_0>0$ will be chosen later. 
We have $T\in \mathbb{R}$ with $T=\infty$ the future singularity
and $T=-\infty$ the future event horizon, so the coordinate $T$ covers only the future FRW region. We have $R\in(-\infty,
R_{\partial})$ where  $R=-\infty$ is the event horizon and $R_{\partial}$ the conformal boundary. Define now Kruskal
coordinates
\begin{equation}
\begin{aligned}
U = \begin{cases}
    e^{T(t)-\rho} & \text{Region II} \\
    -e^{-\zeta + R(\xi) } & \text{Region I} 
\end{cases}
\qquad 
V = 
\begin{cases}
    e^{T(t)+\rho} & \text{Region II} \\
    e^{\zeta + R(\xi) } & \text{Region I} 
\end{cases}.
\end{aligned}
\end{equation}
where we now take Region II temporarily to mean the future part only.
This gives
\begin{align}
    T(U, V) &=  \frac{ 1 }{ 2 }\log(U V), & \rho(U, V) &=  \frac{ 1 }{ 2 }\log(V/U), & U&>0, \\
    \zeta (U, V) &=  \frac{ 1 }{ 2 }\log(-V/U), & R(U, V) &=  \frac{ 1 }{ 2 }\log(-VU),  & U&<0.
\end{align}
The future event horizon is now at $U=0$ and the past event horizon at $V=0$.
The octant $V\geq U \geq 0$ covers the future part of of region I, with $\rho=0$ at $U=V$. 
The singularity lies at $(U>0, V=\infty)$. The conformal boundary is at $VU = -e^{2R_{\partial}}$, 
and region I is covered by the regions $U\leq 0$ and $0\leq V \leq -\frac{ e^{2R_{\partial}} }{ U }$. 

Let us now define the functions
\begin{align}
    \hat{t}(X) = T^{-1}\left( \frac{ 1 }{ 2 }\log X \right), \qquad
    \hat{\xi}(X) = R^{-1}\left( \frac{ 1 }{ 2 }\log(-X) \right),
\end{align}
so that $t(U, V)=\hat{t}(UV)$ and $\xi(U, V) = \hat{\xi}(UV)$. The domain of
$\hat{t}$ is $X \in (0, \infty)$. Since $R\in (-\infty, R_{\rm max})$, we have that the domain of $\hat{\xi}$ is $X\in(-e^{2 R_{\rm
max}}, 0)$.
Finally changing the coordinates, we find that the metric is 
\begin{equation}\label{eq:FRWkruskal}
    \dd s^2 = b(UV)\left[ - \dd U \dd V + \left( \frac{ V-U }{ 2 }\right)^2 \dd \Omega^2 \right],
\end{equation}
where
\begin{equation}\label{eq:bfunc}
    b(UV) = \begin{cases}
    \frac{ \hat{a}(\hat{t}(UV))^2 }{ \sqrt{U^2 V^2} }, & U>0, V>0, \\
        \frac{ a(\hat{\xi}(UV))^2 }{ \sqrt{U^2 V^2} }, & U<0, V>0.
    \end{cases}
\end{equation}
Note that the inverse functions $T^{-1}$ and $R^{-1}$ must be computed numerically, and so the same is also true of
$b$.

In order for $b(X)$ to be continous at $X=0$ we need to choose $\xi_0, t_0$ appropriately.
For small arguments we have
\begin{equation}
\begin{aligned}
    \frac{ \log(-UV) }{ 2 } &= R(\xi) = \log\left(\frac{ \xi }{ 4 }\right) - \frac{ 1 }{ 3 }\log f(\xi_0) +
    \mathcal{O}(\xi^2), \\
    \frac{ \log UV }{ 2 } &= T(t) = \log\left(\frac{ t }{ 4 }\right) - \frac{ 1 }{ 3 }\log \hat{f}(t_0) +
    \mathcal{O}(t^2), \\
\end{aligned}
\end{equation}
which at small $t$ and $\xi$ gives the relation
\begin{equation}
\begin{aligned}
    \xi &= 4 f(\xi_0)^{1/3}\sqrt{-U V}, \\
    t &= 4 \hat{f}(t_0)^{1/3}\sqrt{U V}. \\
\end{aligned}
\end{equation}
Now, near the horizon we have that $\hat{a}(t)=t+\ldots$ and $a(\xi) = \xi + \ldots$, so the function $b$ near the horizon reads
\begin{equation}
\begin{aligned}
    b(UV) = \begin{cases}
    16 \hat{f}(t_0)^{2/3} + \ldots  \\
    16 f(\xi_0)^{2/3} + \ldots  \\
\end{cases}.
\end{aligned}
\end{equation}
Thus, continuity of $b$ requires $t_0$ and $\xi_0$ to be related by
\begin{equation}
    \begin{aligned}\label{eq:fmatching}
    f(\xi_0) = \hat{f}(t_0).
\end{aligned}
\end{equation}

\subsubsection*{The conformal diagram}
The metric \eqref{eq:FRWkruskal} is just conformal to Minkowski, and so drawing the conformal diagram is just as for
Minkowski, with two exceptions:
\begin{itemize}
    \item In the region of negative $U$, $-UV \geq e^{2 R_{\partial}}$ is excised since it lies beyond the
        conformal boundary.
    \item The part which corresponds to null infinity in Minkowski (and which is not in the excised region) is here
        instead a singularity at a finite proper distance.
\end{itemize}
From these observations, we easily find Fig.~\ref{fig:conformaldiagram}. 
In our representation have chosen $t_0=1$ and rescaled the null coordinates by a
convenient overall factor in order to bring the holographic screen closer to the center of the diagram.

\subsubsection*{The holographic screen}
Consider the radial null vectors 
\begin{equation}
\begin{aligned}
    k^a = \frac{ 1 }{ \sqrt{2} \hat{a}(t)} \left[ (\partial_t)^a + \frac{ 1 }{ \hat{a}(t) } (\partial_{\rho})^a \right], 
    \qquad \ell^a  = \frac{ \hat{a}(t) }{ \sqrt{2} }\left[ (\partial_t)^a - \frac{ 1 }{ \hat{a}(t) }(\partial_{\rho})^a
    \right],
\end{aligned}
\end{equation}
which are normalized so that $k\cdot \ell = -1$ and so that $k^a \nabla_a k^b = 0$. Calculating the expansions, we find
\begin{equation}
\begin{aligned}
    \theta_{k} &= \frac{ \sqrt{2} }{ \hat{a}(t)^2 }\left(\hat{a}'(t) + \frac{ 1 }{ \tanh \rho }\right) \\
    \theta_{\ell} &= - \sqrt{2}\left(-\hat{a}'(t) + \frac{ 1 }{ \tanh \rho }\right) \\
\end{aligned}
\end{equation}
For times where $\hat{a}'(t) < -1$, we have marginally trapped surfaces at
\begin{equation}
\begin{aligned}
    \rho(t) &= \arctanh\left( -\frac{ 1 }{ \hat{a}'(t) }\right). 
\end{aligned}
\end{equation}
The unnormalized tangents to the screen that are orthogonal to the marginally trapped leaves are
\begin{equation}
\begin{aligned}
    \eta^a = (\partial_t)^a + \rho'(t) (\partial_{\rho})^a = (\partial_t)^a + \frac{ \hat{a}''(t) }{ 1-\hat{a}'(t)^2
    }(\partial_{\rho})^a.
\end{aligned}
\end{equation}
Its norm is given by
\begin{equation}
\begin{aligned}
    \eta^2 = -1 + \frac{ \hat{a}(t)^2 \hat{a}''(t)^2 }{ (\hat{a}'(t)^2 - 1)^2 },
\end{aligned}
\end{equation}
which starts out being positive and then transitions to negative at late times.

\subsubsection*{Timelike geodesics must cross the horizon}
Consider a timelike geodesic in region II, which by spherical symmetry can be taken to lie at $\theta=\pi/2$ without loss of
generality. The effective Lagrangian for a geodesic is 
\begin{equation}
\begin{aligned}
    \mathcal{L} = g_{\mu\nu}u^{\mu}u^{\nu} = \dot{\xi}^{2} - a(\xi)^2 \dot{\zeta}^2 + a(\xi)^2 \cosh^2 \zeta
    \dot{\varphi}^2,
\end{aligned}
\end{equation}
Parametrizing by proper time, so that $\mathcal{L}=-1$, gives
\begin{equation}
\begin{aligned}
    -\dot{\zeta}^2 + \cosh^2 \zeta \dot{\varphi}^2 = - \frac{ 1 + \dot{\xi}^2 }{ a^2  }.
\end{aligned}
\end{equation}
Then the geodesic equation for $\xi$ can then be written
\begin{equation}
\begin{aligned}
    0 &= \ddot{\xi} - a a'\left( -\dot{\zeta}^2 + \cosh^2 \zeta \dot{\varphi}^2\right) = 
    \ddot{\xi} + \frac{a'}{a}(1+ \dot{\xi}^2).
\end{aligned}
\end{equation}
Interestingly the angular momentum makes no presence here, so there is no angular moment barrier between the horizon and
the conformal boundary. Since $a'(\xi)/a(\xi) > 1$ everywhere, this equation effectively describes a particle subject to
friction and a force always pushing in the direction of negative $\xi$. Thus no geodesic can
avoid reaching $\xi=0$. After this, it enters region II, where it is doomed to encounter the singularity in finite
time. Thus, every timelike geodesic ends up in the singularity, and so the singularity is rightfully described as
cosmological.

\subsubsection*{Computing $\dot{\mathcal{C}}_V$}
Let us consider the static cylinder conformal frame. 
Defining $\xi= - \log z$, the metric in the causal wedge becomes
\begin{equation}
\begin{aligned}
    \dd s^2_{I} &= \frac{ 1 }{ z^2 }\left[\dd z^2 + f(z)^2\left(\dd \zeta^2 +
        \cosh^2\zeta \dd \Omega^2 \right)\right] \\
            f(z) &= \frac{ \left[1-z\right]\left[1+c(\sqrt{z}-1)^2 +
            z\right] }{ 2 } = \frac{ 1+c }{ 2 } - c \sqrt{z} + \mathcal{O}(z) .
\end{aligned}
\end{equation}
Next we introduce new coordinates $(w, t)$ through:
\begin{equation}
\begin{aligned}
    \zeta &= 2 \arctanh\left(\tan \frac{ t }{ 2 }\right), \\
    z &= w \frac{ c+1 }{ 2 \cos t }.
\end{aligned}
\end{equation}
To subleading order in $w$, this transforms the metric into Fefferman-Graham
coordinates of the static cylinder:
\begin{equation}
\begin{aligned}
    \dd s^2_{I} &= \frac{ 1 }{ w^2 }\left[\dd w^2 + h(w, t)^2 \left(-\dd t^2 +
    \dd \Omega^2\right)+\mathcal{O}(w) \right], \\
            h(w, t) &= 1 - \sqrt{w} \sqrt{ \frac{ 2c^2 }{ (c+1) \cos t  } },
\end{aligned}
\end{equation}
where $t$ lies in the finite range $|t|< \pi/2$.
Consider now a hypersurface $\Sigma_{t_0}$ with embedding coordinates $(w, t(w), \Omega_i)$,
and where $t(0)=t_0$.
Its volume reads
\begin{equation}
\begin{aligned}
    \vol[\Sigma_{t_0}] = \vol[S^{d-1}] \int \dd w \frac{ h(w, t)^{d-1} }{
    w^{d} }\sqrt{1 - h(w, t) t'(w)^2}.
\end{aligned}
\end{equation}
Expanding near the boundary,
\begin{equation}
\begin{aligned}
    t(w) = t_0 + t_1 w + \mathcal{O}(w^{3/2}),
\end{aligned}
\end{equation}
we find that extremality imposes $t_1=0$.
Consequentially, the divergence of the volume to subleading order is
\begin{equation}
\begin{aligned}
    \vol[\Sigma_{t_0}] &= \vol[S^{d-1}] \int_{\epsilon} \dd w\left(
    w^{-d}-w^{-d+1/2}(d-1)\sqrt{\frac{ 2 c^2 }{ (c+1)\cos t_0}} + \mathcal{O}(w^{-d+1})  \right)
         \\
                       &=  \vol[S^{d-1}]\left[\frac{ 1 }{ (d-1)  }\epsilon^{-d+1} - \frac{ d-1 }{
                d- \frac{ 3 }{ 2 }}\epsilon^{-d + \frac{ 3 }{ 2 }}\sqrt{\frac{ 2 c^2 }{
        (c+1)\cos t_0 }} + \mathcal{O}(\epsilon^{-d+2}) \right].
\end{aligned}
\end{equation}
This implies that the leading order complexity change with cutoff adapted to the static
cylinder is given by \eqref{eq:staticCyl}.

\subsubsection*{Plot of the Cosmological Wormhole}
\begin{figure}[htb]
\centering
\includegraphics[width=0.99\textwidth]{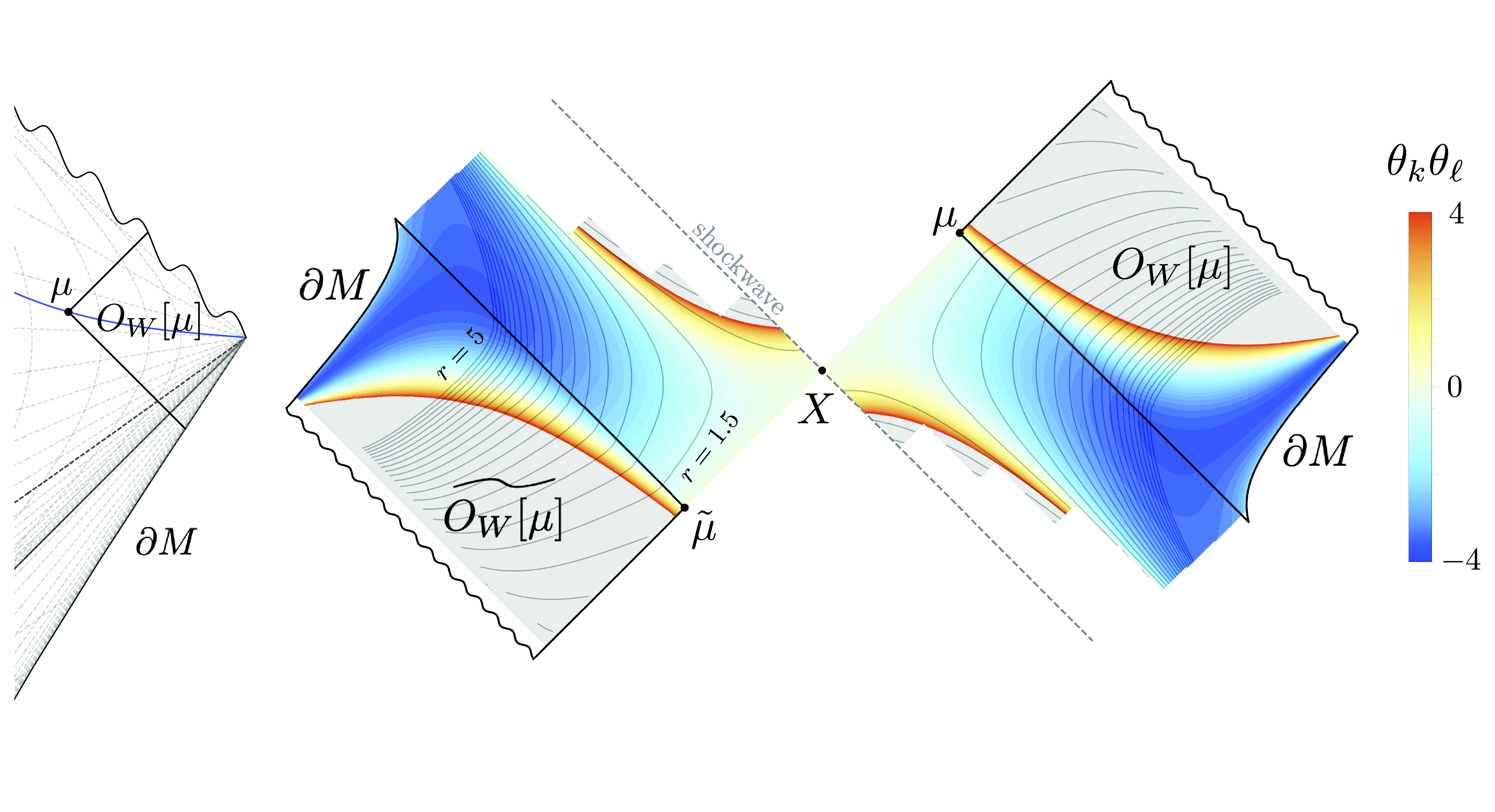}
    \caption{To the left, we have zoomed into the outer wedge $O_W[\mu]$ of a
    marginally trapped surface $\mu$ on the spacelike
    section of the holographic screen in Fig.~\ref{fig:conformaldiagram}. On the
    right, we show the coarse-grained spacetime corresponding to $\mu$ in the
    regions where we have been able to obtain the metric numerically. The black
    contour lines show surfaces of constant area radius, saturating at $r=5$ and
    with spacings of $\delta r \approx 0.2$.
    The colored contours show the product of the null expansions 
    $\theta_{k}\theta_{\ell}$ for
    constant$-r$ surfaces, with
    gray regions corresponding to $\theta_{k}\theta_{\ell}>4$.
    The shockwave passing through the HRT surface $X$ carries
    no null energy, but does source a discontinuity in the inaffinity of
    $\ell^a$, which is the null vector along the direction of the shockwave. The
    quantity $\ell^a \nabla_a \phi$ is discontinuous at $X$. }
\label{fig:cosmowormhole}
\end{figure}

Fig.~\ref{fig:cosmowormhole} shows the domain for which the numerical determination of the coarse grained
spacetime has been carried out, together with data on null expansions and area--radii in the geometry.

\bibliographystyle{jhep}
\bibliography{all}

\end{document}